# Tuning thermoelectric properties of $Sb_2Te_3$-$AgSbTe_2$ nanocomposite thin film - synergy of band engineering and heat transport modulation

[1]Abhishek Ghosh, [2]Khushboo Agarwal, [2]Sergio Gonzalez Munoz [1]Prashant Bisht, [1]Chandan K Vishwakarma, [1]Narinder Kaur, [3]Mujeeb Ahmad, [4]Per Erik Vullum, [4]Branson D.Belle[b], Rajendra Singh[1], [2]O. V. Kolosov, [1,5]Bodh Raj Mehta*

[1] Department of Physics, Indian Institute of Technology Delhi, New Delhi, 110016, India

[2]Department of Physics, Lancaster University, Lancaster, LA14YG, United Kingdom

[3]International Research Centre MagTop, Institute of Physics, Polish Academy of Sciences, Aleja Lotnikow 32/46, PL-02668 Warsaw, Poland

[4]SINTEF Industry, Materials Physics, Forskningsveien 1, NO – 0373, Oslo, Norway

[5]Directorate of Research, Innovation, and Development, Jaypee Institute of Information technology, Noida (U.P.), 201309, India

## Abstract:

The present study demonstrates a large enhancement in the Seebeck coefficient and ultralow thermal conductivity (TE) in $Sb_2Te_3$- $AgSbTe_2$ nanocomposite thin film. The addition of Ag leads to the in-situ formation of $AgSbTe_2$ secondary phase nano-aggregates in the $Sb_2Te_3$ matrix during the growth resulting in a large Seebeck coefficient and reduction of the thermal conductivity. A series of samples with different amounts of minor $AgSbTe_2$ phases are prepared to optimize the TE performance of $Sb_2Te_3$ thin films. Based on the experimental and theoretical evidence, it is concluded that a small concentration of Ag promotes the band flattening and induces a sharp resonate-like state deep inside the valence band of $Sb_2Te_3$, concurrently modifying the density of states (DOS) of the composite sample. In addition, the electrical potential barrier introduced by the band offset between the host TE matrix and the secondary phases promotes strong energy-dependent carrier scattering in the composite sample, which is also responsible for enhanced TE performance. A contemporary approach based on scanning thermal microscopy is performed to experimentally obtain thermal conductivity values of both the in-plane and cross-plane directions, showing a reduced in-plane thermal conductivity value by ~ 58% upon incorporating the $AgSbTe_2$ phase in the $Sb_2Te_3$ matrix. Benefitting from the synergistic manipulation of electrical and thermal transport, a large ZT value of 2.2 is achieved at 375 K. The present study indicates the importance of a combined effect of band structure modification and energy-dependent charge carrier scattering along with reduced thermal conductivity for enhancing TE properties.

## 1. Introduction

Thermoelectric materials and devices are based on a direct conversion of heat into electricity which has long been considered a favourable substitute for the global energy predicament. Direct generation of electrical power from waste heat makes them ideal for microelectronics, wearable devices, automotive, and space technology. [1-4] However, the low efficiency of thermoelectric (TE) materials, especially at low to mid-temperature ranges, has been the primary impediment in replacing the traditional power generation methods. The fundamental challenge to designing efficient thermoelectric stems from conflicting transport parameters comprising TE figure of merit $ZT$, defined as $S^2\sigma T/\kappa$, where $\kappa$ is the total thermal conductivity comprising both electronic ($\kappa_e$) and lattice contributions ($\kappa_l$), $S$ is Seebeck coefficient, and σ is electrical conductivity. Recent advancements in TE have mostly emphasized finding materials with intrinsically low thermal conductivity with moderate Seebeck coefficient. Some of the reported materials with low lattice thermal conductivity, thereby improved ZT, belong to silver chalcogenide-based hybrid binary and multinary compounds systems such as LAST ($AgPb_mSbTe_{2+m}$) [5-9] and TAGS-x ($(GeTe)_x(AgSbTe_2)_{100-x}$),[10-12] which have been widely considered as promising bulk nanostructured TE materials operating at mid-temperature to the high-temperature range. Intrinsically low lattice thermal conductivity values in these systems stem largely from variations of the local interatomic bonding environment between $Ag^+$ and $Sb^{3+}$ cations, resulting in strong phonon scattering.[13] The system LAST-m results in a $ZT$ value of 2.2 at 800 K when m is 10 or 18, while the TAGS-$x$ reaches a $ZT$ of over 1.5 at 800 K when the value of $x$ is 80 or 85.[6, 14] Recent research indicates that $AgSbTe_2$ materials doped with S and Se display an excellent figure of merit of 2.3 at 673 K.[15] Similarly, by utilizing band engineering and inducing stacking defects, p-type $AgSbTe_{2-x}Se_x$ alloys with a ZT value larger than 2.0 are produced. [16] The progress related to the enhancement of $ZT$ value in this class of material owes to the design of non-stoichiometric compositions either by controlling the secondary phases or by introducing lattice anharmonicity by doping, which can intensify the phonon scattering leading to low $\kappa_l$.[17-19]. However, with the phonon thermal conductivity of these materials approaching the minimum theoretical limit (amorphous limit), it is imperative to simultaneously reduce thermal conductivity and enhance the power factor to realize a high *ZT* value for practical applications.

Motivated by the intriguing TE properties of silver chalcogenide, we chose nanostructured hybrid AST ($Sb_2Te_3$- $AgSbTe_2$)-based thin film, which is still missing from the diversity of TE materials. Lower dimensional thermoelectric devices offer numerous advantages over their bulk configuration, including high integration density in compact systems, higher output voltages for smaller temperature differences, and high power densities due to thinner thermocouples and greater heat fluxes. However, the major challenge in these systems is difficulty in accurately determining the transport parameter comprising ZT, especially the thermal conductivity value. Thereby, the present study aims to systematically investigate the electronic and thermal transport properties of nanostructured AST-based thin film to achieve a high ZT value comparable to the bulk counterpart. For this purpose, a controlled concentration of Ag is incorporated in $Sb_2Te_3$ thin film during the film growth, resulting in the formation of the $Sb_2Te_3$-$AgSbTe_2$ hybrid system. Ag-Sb-Te ternary composite thin films with different $Sb_2Te_3$/$AgSbTe_2$ ratios were synthesized, and the dependence of the resulting TE properties on the microstructure has been systematically studied as a function of Ag concentration. A large TE power factor of 7.17 $mW/mK^{-2}$ has been achieved at 435 K in the present study.

In addition, a novel technique of cross-sectional scanning thermal microscope (xSThM) has been adopted that allows for the measurement of anisotropic (in-plane and cross-plane) thermal conductivity in thin layers of materials, thereby overcoming a significant obstacle encountered by conventional thermal conductivity measurement techniques.[20-23] The samples were prepared using the beam exit cross-sectional polishing (BEXP) technique, which uses Ar ions to create close to atomically flat wedge cuts.[24] The SThM measurements conducted on such wedge geometries facilitate the determination of thermal resistance as a function of thickness by considering each measurement point as a sample having a different thickness. These measurements, in conjunction with a simple analytical model (Muzychka-Spiece), help in determining the anisotropic thermal conductivity values ($k_{xy}$ and $k_z$, respectively). A significant reduction in $\kappa$ value is observed in the composite sample, resulting in a high ZT value of 2.2 at 375K, which is among the highest ZT value reported in AST based systems. Moreover, this method has the ability to eliminate the thermal probe–sample interfacial thermal resistance, which otherwise is difficult to determine and often ignored in such multi-interfaced nanocomposites.

The present study is based on experimental and theoretical investigations comprising a series of five nanocomposite samples with different Ag concentrations. The atomic percentage of Ag in the composite sample is denoted by x. The pristine $Sb_2Te_3$ sample ($x$=0) is termed ST, while samples with $x$=0.9, 1.70, 5.90, and 7.00 are termed AST1, AST2, AST3, AST4 respectively.

## 2. Results and Discussions

### 2.1 Structural characterizations

The electronic microstructure and phase modification in these nanocomposites were characterized by a combination of X-ray diffraction (XRD), X-ray photoelectron spectroscopy (XPS), and Transmission electron microscopy (TEM). Fig. 1a shows the XRD spectra for pristine $Sb_2Te_3$ and $Sb_2Te_3$-$AgSbTe_2$ nanocomposite samples grown at a substrate temperature of 523 Kelvin (K). All the samples inherit typical trigonal $Sb_2Te_3$ (space group R3m) structure with prominent diffraction peaks positioned at 28.41$^{o}$, 38.41 $^{o}$, 42.51 $^{o}$, 58.57 $^{o}$,74.81$^{o}$, which correspond to (015), (1010), (0015), (125) crystal planes, respectively, (JCPDS NO-01-071-0393). As the Ag concentration increase, a new peak appears at 29.53 $^{o}$, identified as the (200) lattice plane of the $AgSbTe_2$ phase (JCPDS No-00-015-0540). The co-occurrence of a new secondary phase indicates that the introduction of elemental Ag triggers the growth of the $AgSbTe_2$ phase in the $Sb_2Te_3$ film during the growth process. The diffraction peak intensity of $AgSbTe_2$ gradually increases as the proportion of Ag increases and completely dominates the $Sb_2Te_3$ peaks for higher Ag concentrations. The prominent peaks arising due to the $AgSbTe_2$ phase in the XRD spectra of the composite can be related to the suppression of the crystallinity of $Sb_2Te_3$. From the literature, it is widely accepted that Ag acts as an inhibitor to the crystallization process of $Sb_2Te_3$, leading to a rise in the crystal-

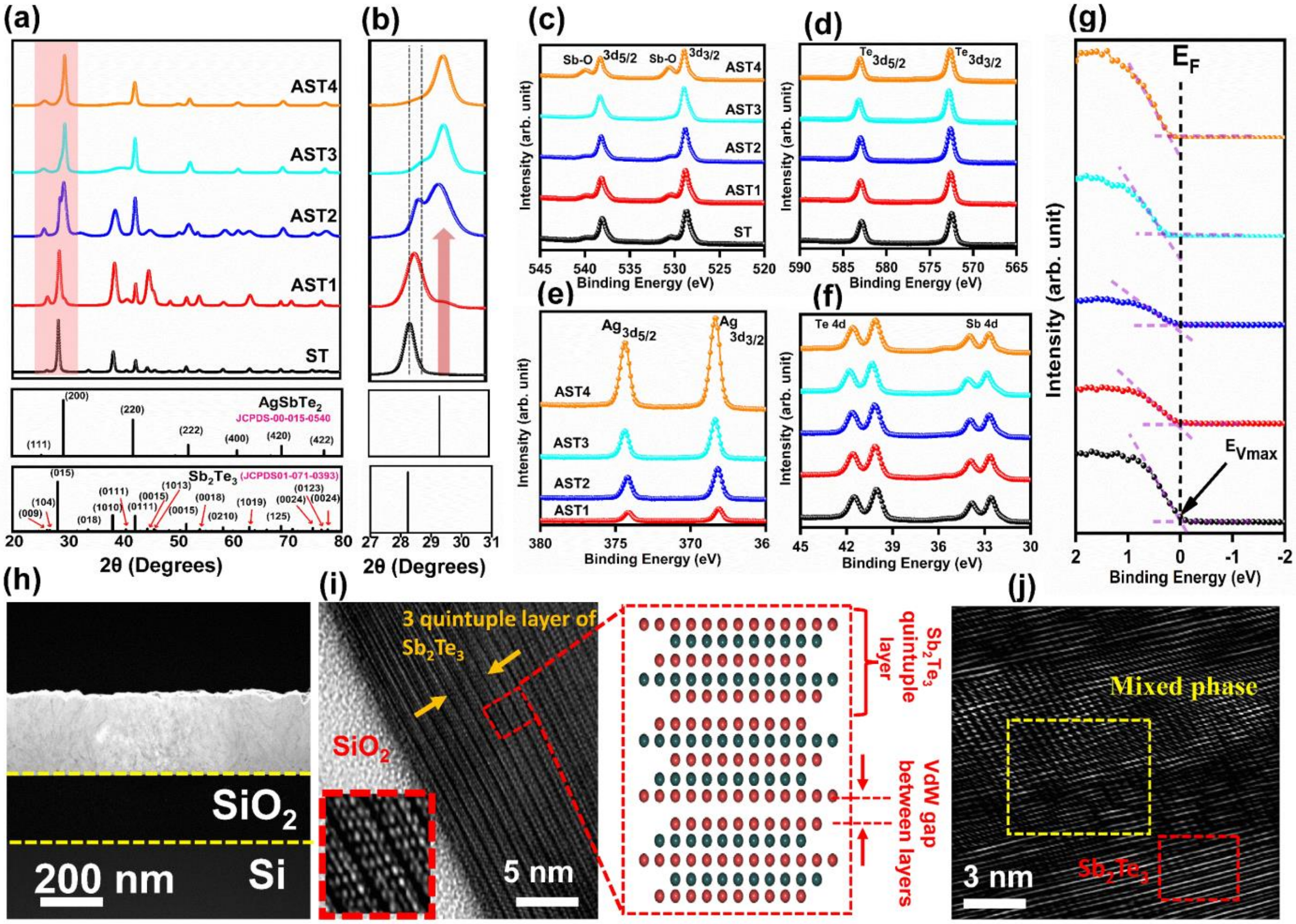

**Figure 1.** (a) θ-2θ x-ray diffraction (XRD) patterns for the as grown films with different Ag composition varying from 0.0 to 7.00 atomic percentage. (b) Coloured square is guide to see the evolution of $AgSbTe_2$ phases which is also highlighted by arrow in the zoomed-in panel to the right. The shift in peak positions of $Sb_2Te_3$ is also indicated by dotted lines. The corresponding JCPDS data is also plotted for comparison. (c-f) XPS spectra showing the change in binding energy with variation of Ag. (g) The valance band spectra showing the shifting in Fermi level. (h) Low-magnification cross section TEM image of AST5 (i) showing the growth of c-axis oriented $Sb_2Te_3$ layers (Sample ST) on Si/$SiO_2$ substrate. A magnified image of quintuple $Sb_2Te_3$ layers is shown as inset and the corresponding atomic arrangements is presented in the offset (j) modifications of $Sb_2Te_3$ crystal structure and formation of two different phases in sample AST4.

line temperature and a decrease in the crystallization activation energy, which is responsible for weakening in the intensity of $Sb_2Te_3$-related XRD peaks with increase in Ag concentration.[25-27] The detailed crystallization process of silver doping on the crystallization behaviour of $Sb_2Te_3$ is examined by Xu et al., where it is assumed that $AgSbTe_2$ serves as the nucleation centre for the growth of the $Sb_2Te_3$ phase.[25]The ionic bonds between Ag and Te atoms are substantially stronger than the covalent ones shared by Sb and Te atoms. Thus, certain Te atoms are unable to acquire sufficient momentum during growth to acquire their locations in the $Sb_2Te_3$ crystal, thereby impeding the nucleation and crystallization process. Further, to validate our speculation, cross-sectional HRTEM is performed and presented in Fig. 1 h-j. For comparison, two sets of figures corresponding to sample ST and AST4 are demonstrated in Fig.1h and Fig. 1j, respectively, showing the modifications in the crystal structure due to Ag incorporation. The (0 0 0 1) oriented parallel lattice fringes of the $Sb_2Te_3$ with a periodicity of five lattice planes can be easily identified in the pristine sample (ST). Three $Sb_2Te_3$ quintuple layers separated by a VdW gap are visible in the magnified image, which is also presented as an inset. In contrast to the pure sample, the long-range crystal structure of the parent $Sb_2Te_3$ phase is disrupted in AST4, and, as shown in Fig. 1j, the majority of it exists as a mixed phase. Moreover, some amorphous phase is observed in AST4, which may be attributable to the less crystalline $Sb_2Te_3$ constituent. (Fig. 9d). Furthermore, almost all peak positions shift slightly towards a higher 2θ value than the pristine sample, suggesting that the unit cell goes through a systematic contraction in the lattice parameter. This type of shifting is also observed in previous reports where $Sb_2Te_3$ was doped with Al or Cr.[28, 29] The reduction in lattice parameters agrees with the concept of substitutional doping, as interstitial doping is usually accompanied by an increase in the lattice parameter and unit cell volume. Among the possible doping sites, the substitution of Ag at Sb sites is more probable than Te, as the electronegativity value of Ag (1.93) is smaller than that of Te (2.1) compared to Sb (2.05).[30] Moreover, the covalent radius of Ag (0.134 nm) is smaller than Sb (0.140 nm); when Ag atoms substitute some Sb atoms, the crystal lattice constants will decrease according to interplanar spacing values and reflect the change in unit cell parameters.[31] Also, in the case of substitutional doping, the formation energy is sufficiently lower compared to the interstitial one, and the Ag substitution for the Sb atom results in the most stable configuration.[27] Clearly, two different growth mechanisms are occurring in Ag-modified samples during the formation of the nanocomposite film. Firstly, Ag atoms replace some Sb/Te atoms, resulting in doping in the $Sb_2Te_3$ phase. In addition, the doping with Ag triggered the formation of a new secondary phase, $AgSbTe_2$. The emergence of $AgSbTe_2$ suggests that the introduction of Ag in $Sb_2Te_3$ is likely to form [$Ag_{0.5}$ -$Sb_{0.5}$] - Te- [$Ag_{0.5}$-$Sb_{0.5}$]- Te layered structure as shown in Fig. 3a, which corresponds to the trigonal system of $AgSbTe_2$. However, it is also possible to form the rock salt structure of $AgSbTe_2$ as experimentally investigated in several reports.[32, 33]

Further, Raman measurements (supporting information Fig. S1) are performed to corroborate the formation of the nanocomposite, as XRD alone is unable to differentiate between the two distinct phases comprising the nanocomposite. The Raman measurements unequivocally show the vibrational peaks corresponding to $Sb_2Te_3$ in all the samples, providing evidence that the $AgSbTe_2$-$Sb_2Te_3$ heterostructure was successfully formed. To investigate the chemical bonding nature of different elements in the nanocomposite samples, XPS measurements were carried out and presented in Fig. 1 c-g. The relative intensities of Ag 3d peaks increase gradually as the Ag content increases from AST2 to AST4. After Ag incorporation, the binding energy (BE) of both Sb and Te is shifted towards higher values confirming a different chemical environment than the pristine sample. Usually, when an atom bonds with another atom possessing a lower electro-negativity, binding energy decreases. Since the electronegativity of Ag (1.9) is less than Sb (2.05) and Te (2.12), when Ag atoms are bonded with Sb and Te atoms in film, it should shift the peaks towards lower binding energy for the Sb and Te elements. However, in the present case, it is observed that the BE for both Sb and Te increases considerably. To interpret the change in BE, XPS valence band spectra are recorded (Fig. 2g), which show a substantial change in the ($E_F$ -$E_{vmax}$), where $E_F$ is the Fermi level, and $E_{vmax}$ is the valence band maxima. Since the Fermi level of material is calibrated as 0 eV in XPS, any $E_F$ change will manifest as a shifting of all the core XPS spectra. So, it is reasonable to assume that the modifications in the binding energy values of the XPS core lines in the present case are due to the combined effect of both chemical shiftings as well as changing the Fermi level due to Ag incorporation. This argument is further justified by KPFM measurement (Supporting Information, Fig. S5), where a systematic shift in the work function value is observed with increasing Ag concentration. Some oxidation peaks are observed in Sb due to the bonding of Te atoms with oxygen being much weaker than Sb, irrespective of Ag incorporation, which can be well explained based on their ionization energy. Elements having smaller ionization energies are more susceptible to losing electrons. So, during the oxidation process, Sb atoms with the highest ionization energy are preferred to offer electrons to O atoms, followed by Te atoms**.** However, no oxide phases are detected by XRD, indicating that oxidation may be limited to the surface layer, and the minority oxide phase at the surface layer is not expected to contribute to the observed thermoelectric properties. Raman measurements were also employed to identify the vibrational modes among Ag, Sb, and Te atoms, as shown in Supporting Information Fig. S1, which strongly supports the above conclusions, providing additional information on the formation of Ag-Sb-Te bonds.

### 2.2 Electronic transport properties

The TE properties of the $Sb_2Te_3$ as a function of varying Ag concentration in nanocomposite samples are investigated over a temperature range of RT to 573 K, with the results presented in Fig. 2. It is interesting to observe that all the transport properties are susceptible to variations in Ag composition. The incorporation of Ag initially increases the conductivity of the composite samples over the measured temperature range. A substantial increase of room temperature electrical conductivity by almost 27% and 78 % is observed in AST1 and AST2, respectively, compared to the pristine sample. Although the conductivity is observed to fall gradually at higher Ag concentrations, remarkable enhancement in the Seebeck coefficient maintains a large power factor in AST3 and AST4 (Fig. 2e) at the higher temperature. The maximum power factor of 4.91 ± 0.4 mW $m^{-1}$ $K^{-2}$ is observed for pristine $Sb_2Te_3$ at 385 K, rising to 7.17± 0.7 mW $m^{-1}$ $K^{-2}$ at 435 K in AST4, which is approximately 46% increment as compared to the $Sb_2Te_3$. A decline in the S at elevated temperature in AST3 and AST4 is observed and can be related to "two carrier conduction" due to the presence of $AgSbTe_2$, which has coexisting electrons and holes at all temperatures.[34, 35] To explain the modifications in the TE transport parameter, the carrier concentration, and mobility values are calculated from Hall measurements (Supporting Information Fig. S2) and presented in Table 1. The Hall coefficient of all the samples shows the domination of positive charge carriers consistent with the Seebeck coefficient. It can be seen that the carrier concentration value initially rises gradually from $1.12 \times 10^{19}$ $cm^{-3}$ in ST to $4.47 \times 10^{19}$ $cm^{-3}$ in AST2, then falls off to $7.71 \times 10^{18}$ $cm^{-3}$ in AST4. The increased carrier density in AST1 and AST2 can be ascribed to the formation of Ag-generated substitutional defects $Ag_{Sb}^{''}$ during the growth process. It is clear that Ag behaves like an acceptor dopant for lower Ag concentration, which accounts for the increased conductivity value.[31, 36] It is interesting to observe that the temperature-dependent electrical conductivity behaviour changes from metallic to semiconducting in samples AST3 and AST4, which could be interpreted as hopping dominated conduction mechanism in the nanocomposite sample. A similar shift from metallic to semiconducting behaviour was also described by Agarwal et al., in which metallic behaviour was recorded for $Bi_2Te_3$, while semiconducting behaviour was observed for $Bi_2Te_3$: Si nanocomposite film. [37]Additionally, samples AST3 and AST4 show a simultaneous decline in the carrier mobility and carrier concentrations, as shown in Fig. 2f. Clearly, the diminished value of electrical mobility and carrier concentration is reflected as a decrement in electrical conductivity. For the majority of thermoelectric materials, the carrier mobility and carrier density are inversely related because of the carrier scattering mechanism. However, in our case, it is observed that the carrier mobility values have an opposite dependence on the carrier concentration. They reduce significantly as the Ag content increases. The significant decline in carrier mobility cannot be explained by considering a simple scenario of carrier filtering or ionized scattering, where a slight negative impact on charge carrier mobility is expected.[38] Neither the anomalous high Seebeck coefficient values in these composite samples can be explained by assuming a simple weighted sum or percolation transport of randomly distributed $AgSbTe_2$ nanodomains within the $Sb_2Te_3$ matrix. In order to understand this significant change in transport parameter, the Seebeck coefficient is plotted as a function of carrier concentration assuming a single parabolic band (SPB) model with the domination of acoustic phonon scattering, and the value of DOS effective mass $m_d^*$ is estimated (Fig. 2d). $m_d^*/m_e$ value of 0.7 is obtained for pristine samples ST at room temperature, which rises to ~1.3 for samples AST3 and AST4. The significant increment of $m_d^*$ in AST3 and AST4 indicate that the enhanced Seebeck coefficient in these samples cannot be attributed solely to the change in carrier concentration, but effective mass also seems to play an additional role as Seebeck coefficient is related to the effective mass by the following relation;

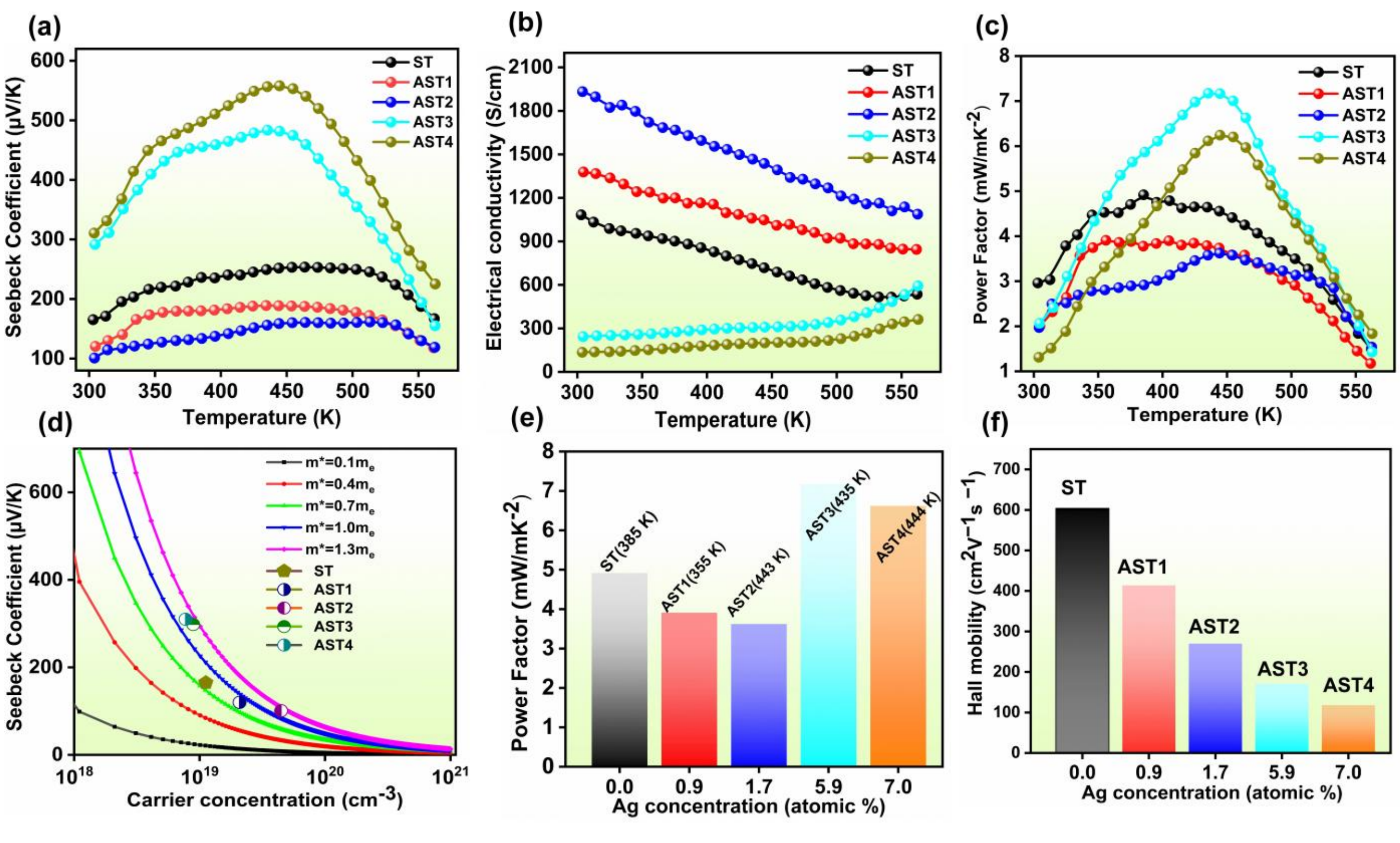

**Figure 2**. Temperature dependence of (a) Seebeck coefficient, (b) electrical conductivity, and (c) power factor for pristine and nanocomposite samples having different concentrations of Ag; (d) variation of Seebeck coefficient as a function of carrier concentration deduced from Piserenko relation; (e) histogram showing the maximum power factor value as a function of Ag concentration; (f) compositional variation of mobility at room temperature. ST, AST1, AST2, AST3, and AST4 represent the $Sb_2Te_3$–$Ag_x$ where $x$ = 0.0, 0.9, 1.70, 5.9 and 7.0, respectively.

**Table 1:** Value of Electrical Conductivity, Carrier Concentration, Hall Mobility, and Seebeck coefficient at room temperature.

| Sample name | Ag Concentration (atomic %) | Electrical conductivity (S/cm) | Carrier concentration ($cm^{-3}$) | Hall mobility ($cm^2V^{-1}s^{-1}$) | Seebeck coefficient (μV/K) | Power factor ($mW\ m^{-1}\ K^{-2}$) |
|---|---|---|---|---|---|---|
| ST | 0.0 | 1083.65 | 1.12E+19 | 603.96 | 165.33 ± 11 | 2.96 ± 0.2 |
| AST1 | 0.9 | 1378.69 | 2.08E+19 | 413.75 | 120.59 ± 8 | 2.00 ± 0.16 |
| AST2 | 1.70 | 1932.79 | 4.47E+19 | 269.90 | 101.02 ± 7 | 1.97 ± 0.15 |
| AST3 | 5.9 | 242.97 | 8.89E+18 | 170.60 | 291.65 ± 20 | 2.06 ± 0.16 |
| AST4 | 7.0 | 135.77 | 7.17E+18 | 118.20 | 310.80 ± 22 | 1.31 ± 0.10 |

$$S = \frac{8\pi^2 k_B^2}{3eh^2} m_d^* T \left(\frac{\pi}{3n}\right)^{2/3} \tag{1}$$

where e is the electronic charge, T is the absolute temperature, $k_B$ and h are the Boltzmann and Planck constants, respectively. [39]DOS effective mass, also referred as the Seebeck effective mass $m^*$, reflects the average density-of-state of the overall band edge and is related to the local band effective mass $m_b^*$ as $m_d^* = N_V^{2/3}\ m_b^*$, where $N_V$ is band degeneracy. $m_b^*$ is inversely proportional to the curvature of electronic bands and can be described by the energy dispersion relationship

$$m_b^* = \hbar^2 \left(\frac{\partial^2 E(k)}{\partial k^2}\right)^{-1} \quad (2)$$

where ħ is reduced Plank constant, k and E(k) represent the wave vector and energy dispersion function, respectively. [40]As indicated by the above equations, the effective mass $m_d^*$ can be increased by increasing the valley degeneracy or by manipulating $m_b^*$ through local distortions in DOS.

### 2.3 Density Functional Theory Calculations

To fully comprehend the change of effective mass and the increased Seebeck coefficient due to Ag inclusion, the band structure and DOS of $Sb_2Te_3$ and $Sb_2Te_3/AgSbTe_2$ heterostructures are calculated using density functional theory and presented in Figure 3. The band structure of the heterostructure shows band folding making it challenging to compare with the band structure of the pure compound. Therefore, the supercell band structures are projected to the primitive unit cell and presented in Fig. 3 c,d,f,g. For a deeper understanding, elemental (Supporting information Fig. S3) and orbital projected band structure (Fig. 3j) are also plotted with the colour of the bands indicating the weight of the local orbitals to the band eigenstates. The analogous feature can also be seen in Fig. 3 h,i,k, where the projected DOS of the majorly contributing states are shown. When compared to pure $Sb_2Te_3$, the heterostructure reveals two key variations in its band structure and DOS. These changes are responsible for the high effective mass of the heterostructure. The first difference between the pure system and the heterostructure is that the bands in the heterostructure are considerably smeared and flattened towards the Fermi level (Fig. 3m). The local crystalline periodicity of $Sb_2Te_3$ is altered by the presence of Ag. Hence, the states exhibit substantial widening and depart from their Bloch character with diminished spectral weight. In accordance with the energy dispersion relation expressed in equation (2), a flat band will result in a high effective mass and, ultimately, a high Seebeck coefficient. In addition, the flattening of the band is related to a widening of the band gap $E_g$, as proposed by the Kane model and characterized by the following equation: [41]

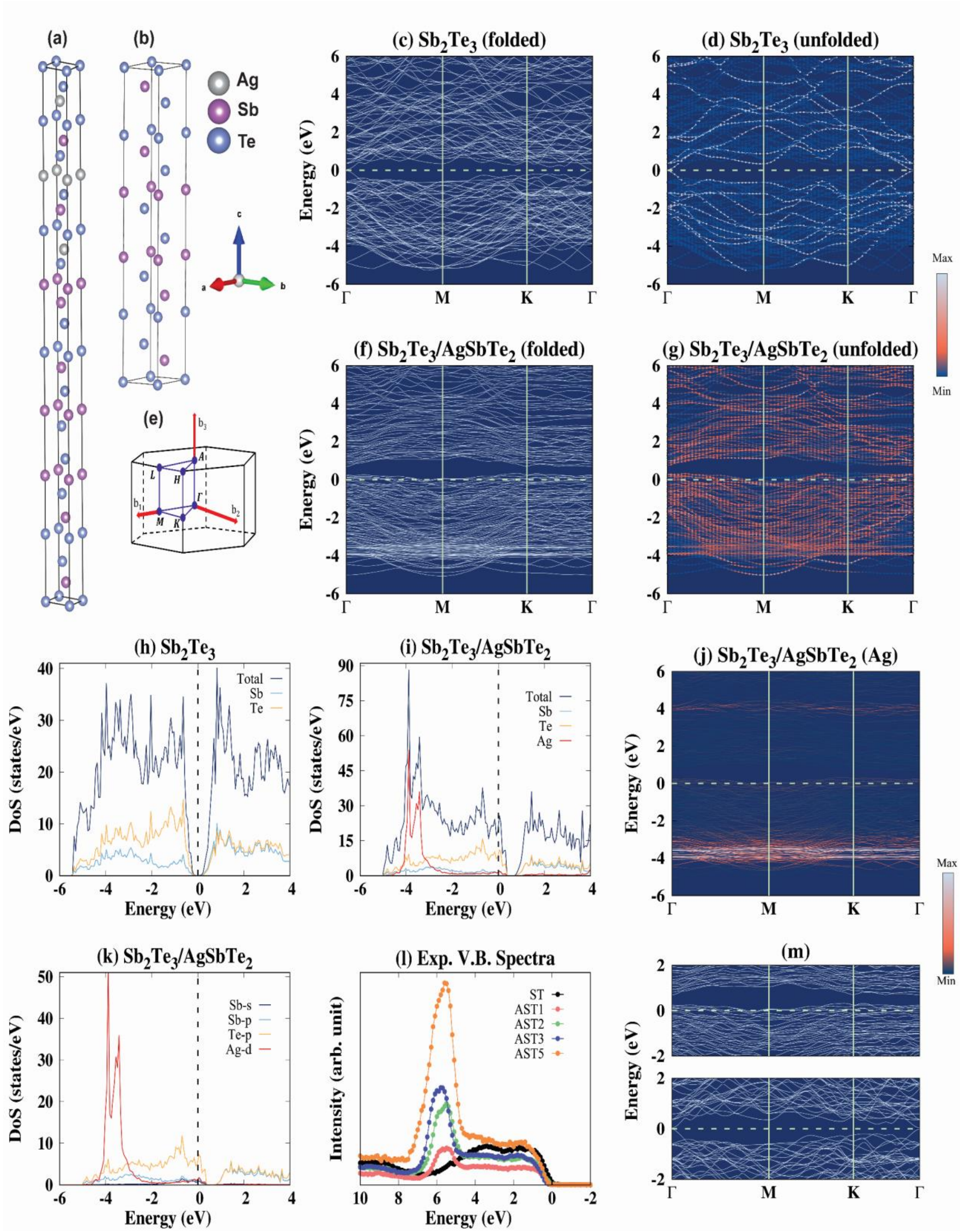

**Figure 3**. a-b Represents the crystal structure corresponding to $Sb_2Te_3/AgSbTe_2$ and $Sb_2Te_3$ respectively. (c-d),(f,g) folded and unfolded band structure of $Sb_2Te_3$ and $Sb_2Te_3/AgSbTe_2$ respectively.(h,i)DOS for the two structure (e) the high symmetry path of $Sb_2Te_3$ (j) Ag contribution to the band structure is presented with colormap. The colour bar indicates the spectral weight at each k points. (g) Brillouin zone path corresponding to the hexagonal crystal structure.(k) partial DOS of the heterostructure (l) experimental valence band spectra for comparison .(m) enlarged images corresponding to (f)(top) and (c)(bottom) to see the evolution of bands around Fermi level.

$$\frac{\hbar^2 k^2}{2m_b^*} = E\left(1 + \frac{E}{E_g}\right) \quad (3)$$

A band opening at the Gamma point in the heterostructure validates the band flattening mechanism, as shown in Figure 3k. To experimentally confirm the band gap elevation, the infrared spectra of pure and composite samples are examined (supporting information Fig. S7), which clearly demonstrates a band gap enlargement in sample AST4. The second notable distinction is a strong localized feature that may be characterized as the Ag 4d states emerge deep within the valence band. This feature has a significant weight in comparison to the background states (Fig. 3l). As the electrons in the d-band lie near the atomic nuclei more than s-electrons, the d-band in a material is relatively localized in energy and can accommodate a larger number of charge carriers than the s-band. As a consequence, the DOS that is contributed by the d band is roughly an order of magnitude more significant when compared to the extended s band. The calculated density of states for heterostructure in Fig. 3 j,k clearly indicates the formation of a heavy band deep inside the valence band, which agrees adequately with the experimentally obtained XPS valence band spectra. The high DOS originating from Ag 4d states cause resonant distortion of the density of states of $Sb_2Te_3$. Notably, the resonant distortion should typically be positioned within a few $K_BT$ in order to increase the Seebeck coefficient. Despite the fact that Ag-induced resonant states are placed far from the Fermi level, which contributes little to the total thermopower, they can produce a scattering mechanism known as resonant scattering. The significant decline in electrical conductivity is analogous to the Tl-doped PbTe system, where the Tl doping generates a hyper-deep resonant level was at approximately 5.5 eV, resulting in higher resistivity than that of Na doping. [42]Additionally, the energy offset between the states significantly decreased in the heterostructure and the Fermi level shift inside the valance band, which could also promote a multi-band transport mechanism with high $N_v$ and high $m_d^*$. As the Seebeck coefficient is directly proportional to effective mass, an increase in $m_d^*$ largely enhances the Seebeck coefficient and thermopower in AST3 and AST4. As the contribution due to $AgSbTe_2$ increases with an increase in impurity concentration, the mechanisms mentioned above are expected to strengthen, and as a consequence, further enhancement in thermopower is observed. A direct manifestation of increased $m_d^*$ is the rapid decrease in the hole mobility as measured by the Hall effect described by the relation $\mu = {}^{1}/_{m_b^*}$. Despite a minute change in the Ag concentrations from 2.60 % to 7.00 %, the mobility drops by more than 60% from 413 $cm^2\ V^{-1}\ s^{-1}$ for AST1 to 118 $cm^2\ V^{-1}\ s^{-1}$ for AST4. The phenomenon of significant mobility reduction is often observed in materials where an increase in the effective mass is involved. As discussed earlier, the effective mass of a band $m_d^*$ is inversely proportional to the curvature of electronic bands; at a constant charge carrier density, dispersive bands with small m* are anticipated to achieve higher electrical mobility than flat bands with large m*, which alleviates carrier mobility reduction.

**Carrier filtering at the interface**

Apart from the increased $m_d^*$ , the formation of an electrostatic potential barrier between the two phases results in energy carrier filtering mechanism, which we believe is partly responsible for improving TE properties. The idea of carrier filtering is to create a potential barrier in the path of charge carriers, which can effectively trap or scatter away the charge carriers having low energy, which are detrimental to the Seebeck coefficient, while high energy carriers can transmit easily without any attenuation. As the Seebeck coefficient is defined as the average energy of carriers around the Fermi level for a given carrier concentration, the Seebeck coefficient is expected to increase significantly. As the heterostructure is formed during the growth, charge transfers occur between the two phases leading to the alignment of the Fermi level, and an equilibrium condition is achieved. The charge transfer at the interface gives rise to a depletion region, which is defined in terms of an electrostatic potential barrier. Kelvin probe force microscopy (Supporting information Fig. S4) and XPS valance band spectra were employed to determine the potential barrier formed at the interface between the two phases. A reference sample with almost exact composition as $AgSbTe_2$ is prepared, and the work function is compared with the pristine $Sb_2Te_3$. The band alignment between $AgSbTe_2$ and $Sb_2Te_3$ is shown in Fig. 4, indicating a barrier height of 0.7 eV between the two phases, sufficient for the carrier filtering effect. It is evident from Fig. 4 that the band offset between the two phases can scatter low-energy charge carriers, whereas high-energy carriers with sufficient energy remain unaffected. This mechanism creates an asymmetry in the carrier relaxation time distribution. As the Seebeck coefficient depends on the energy derivative of the relaxation time, this phenomenon would increase the Seebeck coefficient. [43, 44] The number of interfaces increases with the $AgSbTe_2$ concentration, which in turn intensifies the carrier filtering mechanism. However, due to the carrier filtering mechanism, the number of charge carriers participating to the transport can significantly decrease, which is partly responsible for the decrease in the carrier concentration value and electrical conductivity.

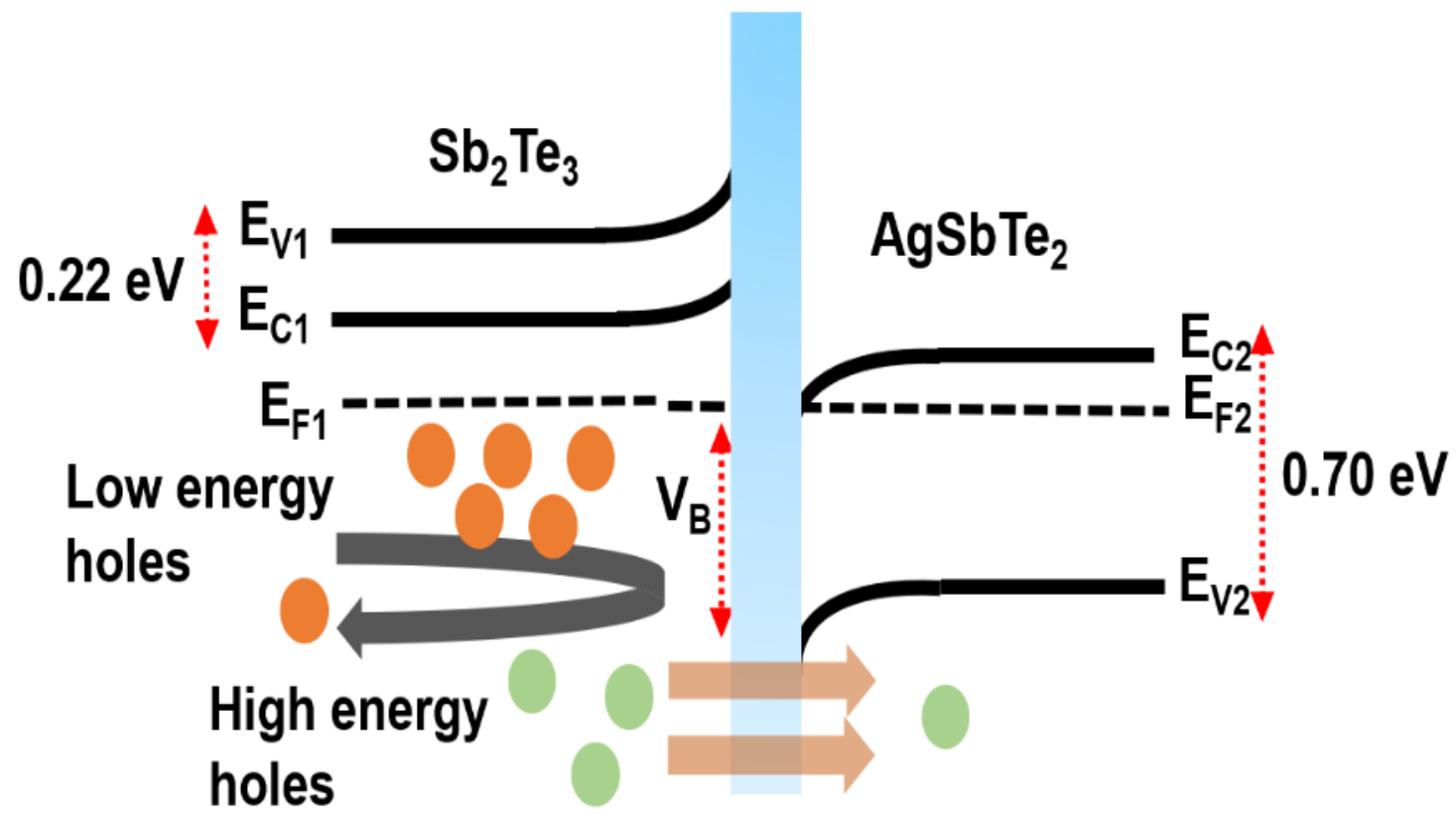

**Figure 4**. Schematic diagram describing the carrier filtering mechanism at the interface between $Sb_2Te_3$ and $AgSbTe_2$

### 2.4 Thermal transport properties

Further, to determine the values of ZT, it is imperative to measure the values of thermal conductivity. Therefore, a novel technique of xSThM (Fig. 5a) was employed to measure the anisotropic thermal conductivity values of such multi-interfaced material. The samples were prepared using the beam exit cross-sectional polishing (BEXP) technique, which uses Ar ions to create close to atomically flat wedge cuts. The SThM measurements conducted on such wedge geometries facilitate the determination of thermal resistance as a function of thickness by considering each measurement point as a sample having a different thickness. These measurements, in conjunction with a simple analytical model

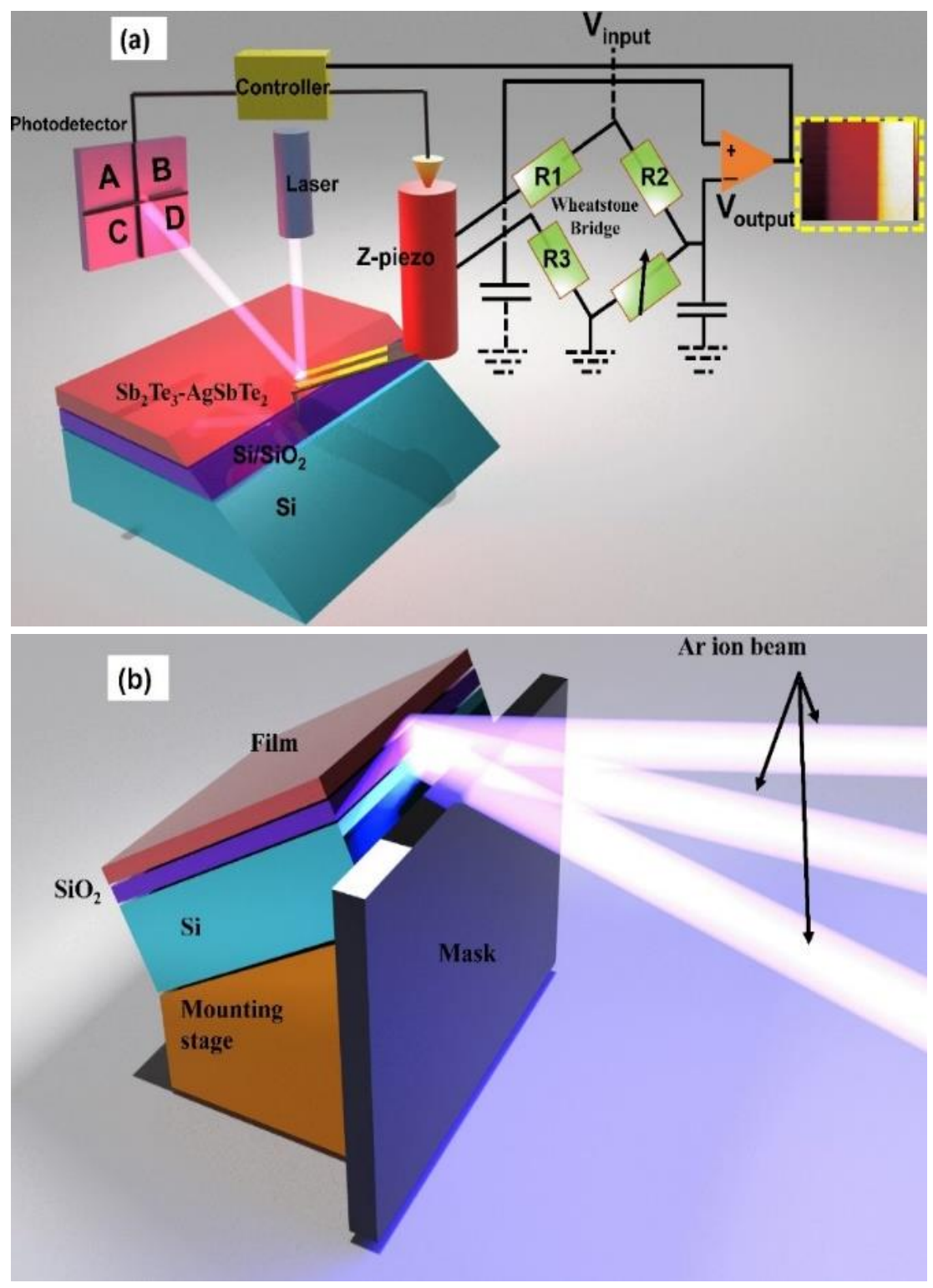

**Figure 5.** Schematic diagram describing the (a) SThM set up used for this study (b) exit cross-sectional polishing (BEXP) technique for sample preparation.

(Muzychka-Spiece), help in determining the anisotropic thermal conductivity values ($k_{xy}$ and $k_z$, respectively). The thermal conductivity measurements were validated using standard photoresist material SU-8 which shows isotropic thermal transport with thermal conductivity values ~ 0.23 W/(mK), matching well with the reported values of 0.2 W/(mK). For xSThM, the cross-sectional samples were prepared using the BEXP technique, as described in the experimental section and also presented in Fig. 5b.[21,22,24] The cross-sectional samples prepared by the BEXP technique were then investigated for SThM measurements under high vacuum conditions (~ $4X10^{-6}$ Torr). Fig. 6 shows the topographic and thermal images at the film-substrate interface of AST1, AST2, AST3, AST4 samples. The variation of thickness and contact voltage ($V_C$) with respect to position curves were acquired from Fig. 6 (a, b), (d,e), (g,h), (j,k) and are plotted in figure 6(c,f,i,l). Due to the lower difference in the height of subsequent layers, it is difficult to identify different materials from the topography images. However, the interface between substrate and the film can be easily isolated from the thermal images due to the difference in thermal properties. It is

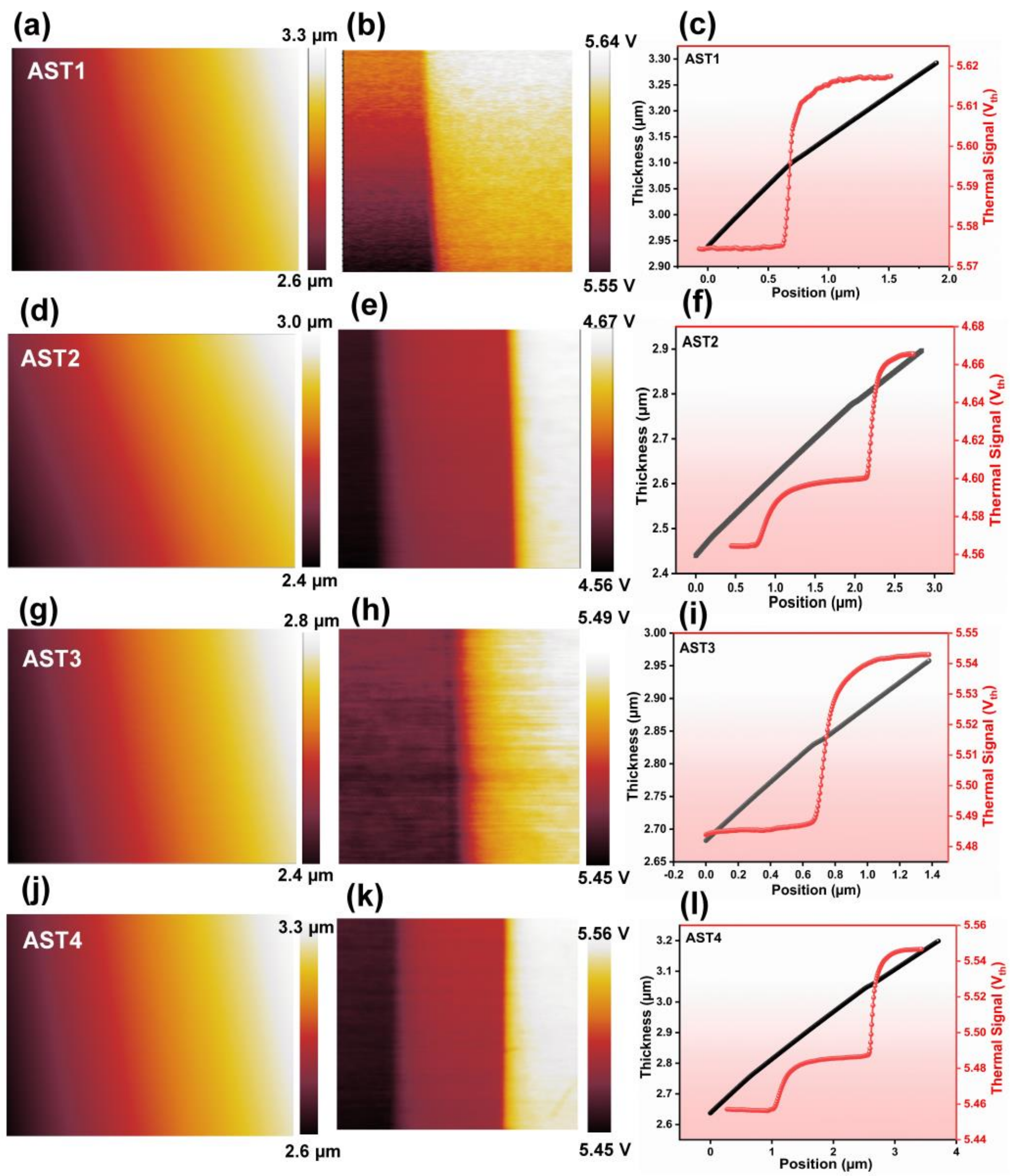

**Figure 6** (a, b), (d,e), (g,h), (j,k) represents the topographical and thermal image for substrate-film interface for AST1, AST2, AST3 and AST4; (c,f,i,l) shows the position dependence of thickness and contact voltage $V_C$ for the same samples determined from the topographical and thermal images. $V_C$ is proportional to the probe temperature at fixed heating power applied to the probe.

worth noting that the position ($x$) is expanded with respect to the actual thicknesses by approximately five times due to the angle of the cut. This can be attributed to the fact that Si, $SiO_2$, and sample have different mechanical properties, and the Ar ions etching rate is slightly different, even at glancing angles, resulting in a slight change of angle at the interface. As each position corresponds to a particular thickness (topography image) and its corresponding contact voltage (thermal image), which can be combined to obtain data of contact voltage $V_c$ as a function of thickness for Si/$SiO_2$ and $SiO_2$/film. $V_c$ represents xSThM "thermal signal" proportional to the probe temperature at fixed heat applied to the probe. After that, the value of thermal resistance ($R_X$) can be evaluated using the below equation

$$R_X(t) = \frac{V_{nc}.R_p}{V_{nc} - V_c(t)} \tag{4}$$

Where, $V_{nc}$ is the non-contact "thermal signal" voltage when the tip is not in contact with the sample surface, Rp is the probe thermal resistance, and $V_c$ is the contact voltage acquired from the thermal images as a function of thickness.[45, 46] Further, to quantify the thermal conductivity of the multi-interfaced material ($Sb_2Te_3$-$AgSbTe_2$), $R_X$ can be defined as the sum of the total probe-sample contact thermal resistance ($R_C$) and the total spreading resistance ($R_S$) within the sample,

$$R_X(t) = R_S(t) + R_C \tag{5}$$

Assuming that , [22] $R_C$ does not vary as a function of thickness; we can use an isotropic model for Si/$SiO_2$ interface, defining $R_S$ as

$$R_s(t) = \frac{1}{\pi\kappa_{tl}a}\int_0^\infty \left[\frac{1 + Kexp\left(-\frac{2\xi\left(t + (r_{int} * \kappa_{tl})\right)}{a}\right)}{1 - Kexp\left(-\frac{2\xi\left(t + (r_{int} * \kappa_{tl})\right)}{a}\right)}\right] J_1(\xi)\sin(\xi)\frac{d\xi}{\xi^2} \tag{6}$$

where $a$ is the radius of the tip, $r_{int}$ is the interfacial thermal resistance per unit area, $t$ is the thickness of the layer, $\xi$ is the integration variable, $J_1$ is the first-order Bessel function, and $K$ is defined as

$$K = \left[\frac{1-\left(\frac{\kappa_S}{\kappa_{tl}}\right)}{1+\left(\frac{\kappa_S}{\kappa_{tl}}\right)}\right] \tag{7}$$

$$R_X(t) - R_C = R_S(t) \tag{8}$$

where, $\kappa_S$ and $\kappa_{tl}$ are the thermal conductivity of the substrate and top layer, respectively. Now, rearranging the equation, we get

At t = 0

$$R_X(0) - R_C = R_S(0) \tag{9}$$

Subtracting equation (9) from equation (8), we get:

$$R_X(t) - R_C - R_X(0) + R_C = R_S(t) - R_S(0) \tag{10}$$

Therefore, equation (10) can be rewritten by substituting the expression for $R_S$ from equation (6), as

$$R_X(t) - R_X(0)$$
$$= \frac{1}{\pi\kappa_{tl}a}\int_0^\infty \left[\frac{1+Kexp\left(-\frac{2\xi\left(t+(r_{int}*\kappa_{tl})\right)}{a}\right)}{1-Kexp\left(-\frac{2\xi\left(t+(r_{int}*\kappa_{tl})\right)}{a}\right)}\right] J_1(\xi)\sin(\xi)\frac{d\xi}{\xi^2}$$
$$- \frac{1}{\pi\kappa_{tl}a}\int_0^\infty \left[\frac{1+Kexp\left(-\frac{2\xi(r_{int}*\kappa_{tl})}{a}\right)}{1-Kexp\left(-\frac{2\xi(r_{int}*\kappa_{tl})}{a}\right)}\right] J_1(\xi)\sin(\xi)\frac{d\xi}{\xi^2} \tag{11}$$

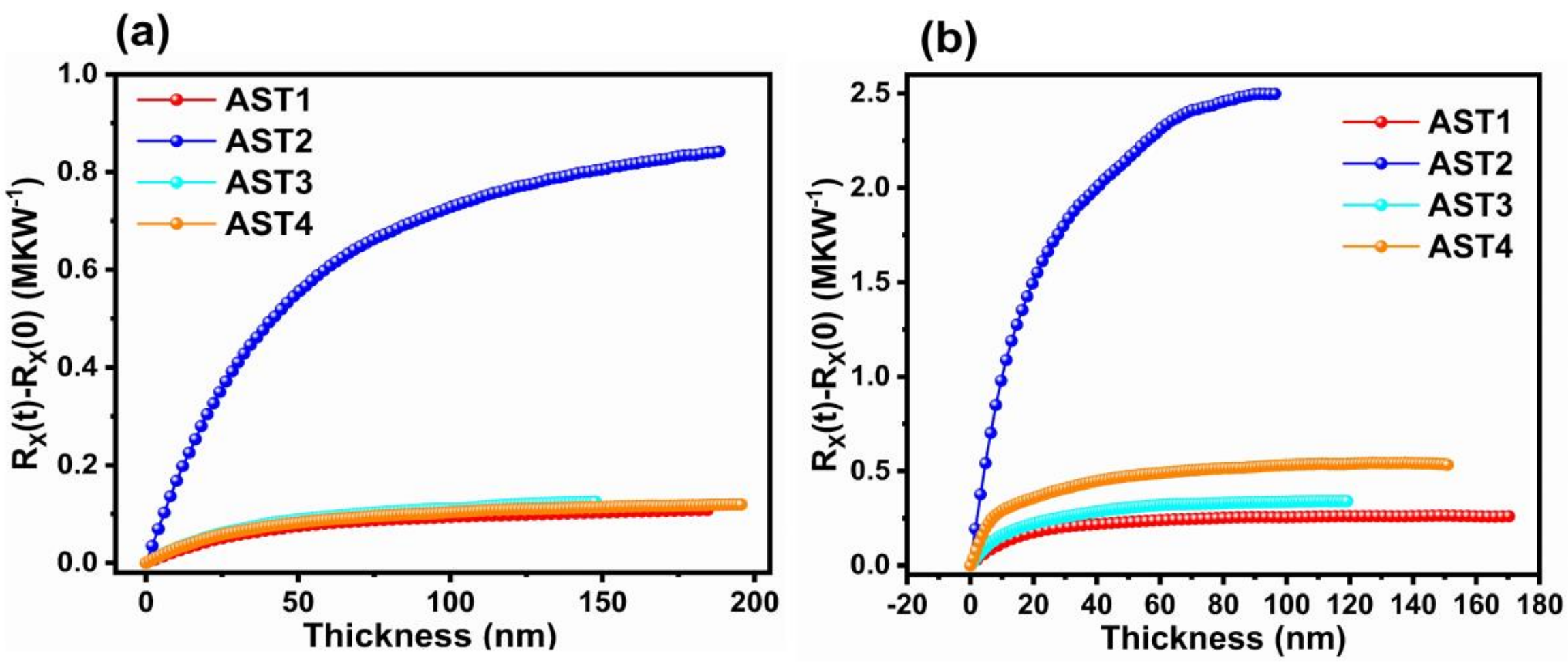

**Figure 7** Contact voltage $A_c$ vs thickness plots for (a) Si/$SiO_2$ interface and (b) $SiO_2$/Sample interface for all the samples.

Fig. 7 (a, b) shows the experimental data of $R_X(t)$ - $R_X(0)$ vs. thickness plots for Si/$SiO_2$ and $SiO_2$/sample interface, respectively, for all the samples. It is worth noting here that Fig. 7 (a, b) has been plotted after shifting the data by a factor of 2a×sin(α) to eliminate the boundary points occurring due to the different thermal resistance values between $SiO_2$ and sample at transition points at the interface. Further, a scaling factor ($c_e$) was also introduced (as a fitting parameter) in equation (11) as the experimental data. The $c_e$ is effectively a ratio in the thermal response of the xSThM thermal sensor at the same heat transferred when the sensor is self-heated and when the heat is transferred to the sample. Thus, equation (11) can be rewritten as

$$c_e[R_X(t) - R_X(0)] = \frac{1}{\pi\kappa_{tl}a}\int_0^\infty \left[\frac{1 + Kexp\left(-\frac{2\xi\left(t + (r_{int} * \kappa_{tl})\right)}{a}\right)}{1 - Kexp\left(-\frac{2\xi\left(t + (r_{int} * \kappa_{tl})\right)}{a}\right)}\right] J_1(\xi)\sin(\xi)\frac{d\xi}{\xi^2} - \frac{1}{\pi\kappa_{tl}a}\int_0^\infty \left[\frac{1 + Kexp\left(-\frac{2\xi(r_{int} * \kappa_{tl})}{a}\right)}{1 - Kexp\left(-\frac{2\xi(r_{int} * \kappa_{tl})}{a}\right)}\right] J_1(\xi)\sin(\xi)\frac{d\xi}{\xi^2} \tag{12}$$

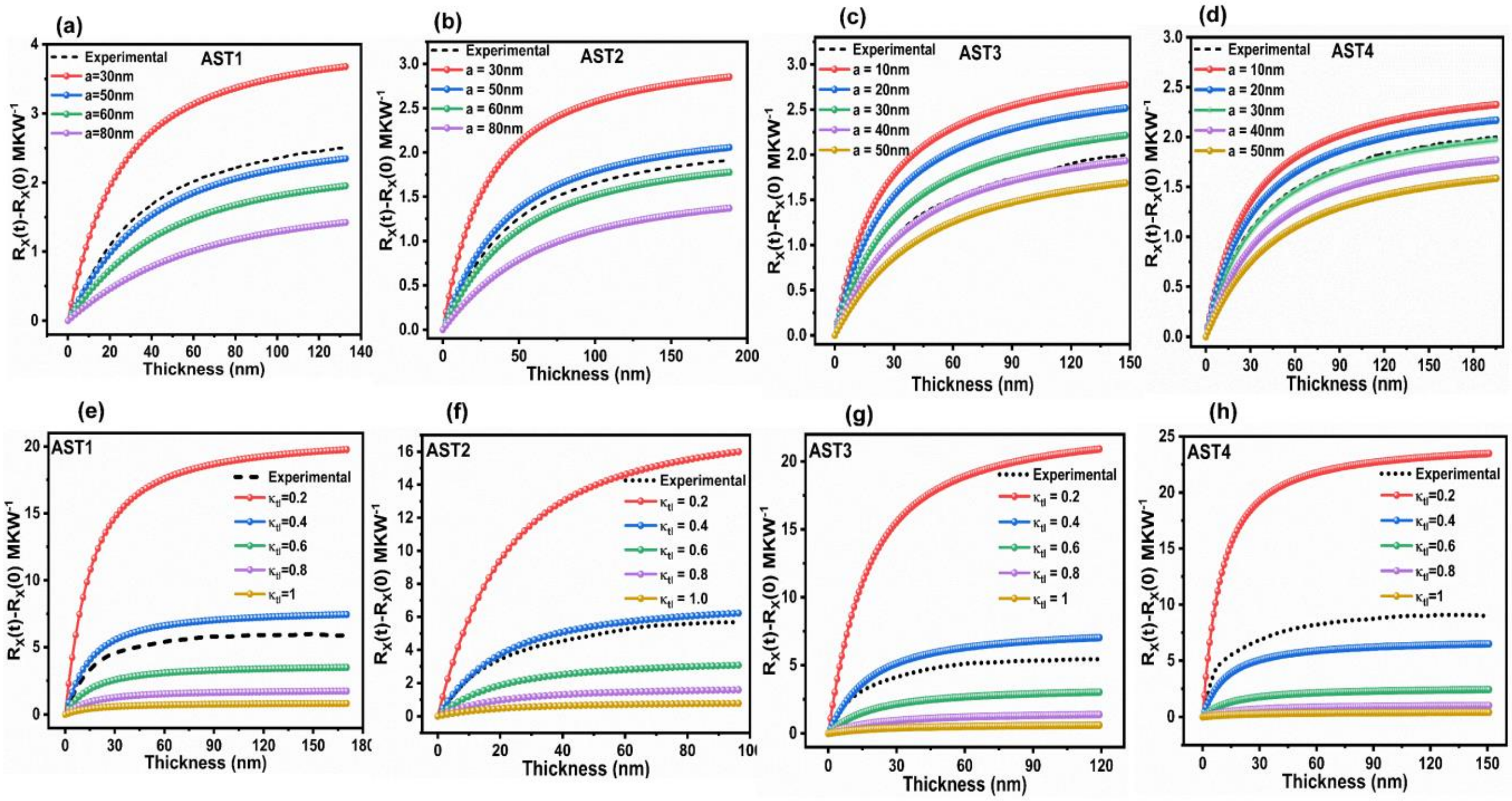

**Figure 8** (a-d) Simulated curves (shown with solid lines) using equation (12) plotted along with the experimental curves (shown with dotted lines) for $Si/SiO_2$ interface needed for estimating the approximate range of parameters $a$ and $c_e$.. (e-h) Analytical curves (shown with solid lines) using equation (12) (for anisotropic model) were plotted along with the experimental curves (shown with dotted lines) of $SiO_2$/Sample interface for estimating the ranges of $\kappa_{tl}$ and $a_r$.

Before fitting the data into the analytical model given by Muzychka-Spiece, the analytical curves using equation (12) were plotted along with the experimental data (Fig. 8) to determine the ranges of the unknown parameters. Therefore, equation (12) (isotropic model) was used to plot analytical curves for $Si/SiO_2$ interface, wherein Si acting as the substrate layer ($\kappa_S = 130$ $Wm^{-1}K^{-1}$) and $SiO_2$ is treated as the top layer ($\kappa_{tl} = 1.4$ $Wm^{-1}K^{-1}$). Thus, considering the respective values of $\kappa_S$ and $\kappa_{tl}$, ranges for the values of $a$ and $c_e$ were determined (Fig. 8 e-h). It is worth noting that the ranges of $r_{int}$ for the $Si/SiO_2$ interface are well-known in the literature and can be used further while fitting the experimental curves in the analytical model. These analytical curves were not only helpful in setting the ranges for the unknown parameters but also gave a useful insight into the effect of variation of these parameters on the nature of the curves. The ranges obtained from the analytical curves were then used in a custom-built Matlab program to obtain the values of $a$ and $c_e$ (from $Si/SiO_2$ interface thermal signal curves). The values of $a$ and $c_e$ determined from the fitting (Table 2) were used to plot analytical curves (shown by dotted line) along with the experimental data of the $Si/SiO_2$ and $SiO_2$/Sample interface for AST1, AST2, AST3, and AST4 samples (shown by solid lines) and are presented in supporting information Fig. S6. The goodness of fit (adjusted R-square > 0.999) clearly indicates that for such kind of experimental data, the respective combinations of values of $a$ and $c_e$ are both unique and accurate. Further, to obtain the ranges of $k_{xy}$ and $k_z$, equation (12) can be rewritten for an anisotropic model (used for $SiO_2$/sample interface), wherein $t$ and $\kappa_{tl}$ were replaced according to the following equations

$$t = t * \sqrt{a_r} \tag{13}$$

$$\kappa_{tl} = \sqrt{k_z * k_{xy}} \tag{14}$$

**Table 2:** Values of $a$ and $c_e$ determined from $Si/SiO_2$ interface and $k_{xy}$ and $k_z$ determined from $SiO_2$/sample interface. ZT values are calculated in the in-plane direction at room temperature. From the value of thermal conductivity obtained at RT, The thermal conductivity value of $Sb_2Te_3$ is taken from Ref. [52]

| Sample Name | $Si/SiO_2$ interface | | | $SiO_2$/Sample interface | | | | $S^2\sigma$ at RT (mW m$^{-1}$ K$^{-2}$) | $S^2\sigma$ at 395K (mW m$^{-1}$ K$^{-2}$) | ZT (RT) | ZT (375K) |
|---|---|---|---|---|---|---|---|---|---|---|---|
| | a (nm) | $c_e$ | $r_{int}\times10^{-9}$ (Km$^2$W$^{-1}$) | $\kappa_{tl}$ (Wm$^{-1}$K$^{-1}$) | $\kappa_{xy}$ (Wm$^{-1}$K$^{-1}$) | $\kappa_z$ (Wm$^{-1}$K$^{-1}$) | $r_{int} * 10^{-9}$ (Km$^2$W$^{-1}$) | | | | |
| ST | - | - | - | - | 1.84 | - | - | 2.96± 0.23 | 4.75 | 0.48 | 0.95 |
| AST1 | 46 | 13.3 | 6.4 | 0.45 | 0.92 | 0.22 | 17.0 | 2.00 ± 0.16 | 3.83 | 0.65 | 1.58 |
| AST2 | 54 | 2.3 | 12.1 | 0.42 | 0.76 | 0.24 | 12.4 | 1.97 ± 0.15 | 3.01 | 0.77 | 1.43 |
| AST3 | 38 | 15.9 | 16.6 | 0.46 | 0.94 | 0.22 | 28.2 | 2.06 ± 0.16 | 6.10 | 0.65 | 2.26 |
| AST4 | 28 | 16.7 | 20.2 | 0.34 | 0.80 | 0.14 | 49.1 | 1.31 ± 0.10 | 4.67 | 0.49 | 1.85 |

Similarly, for the $SiO_2$/sample interface, the analytical curves were initially plotted (Fig. 8 e-h) and analysed to obtain the ranges for the values of '$\kappa_{tl}$' and anisotropic ratios ('$a_r$' = $k_{xy}/k_z$). The ranges obtained from analytical curves were then used as input in the custom build Matlab program to determine the exact values of $k_{xy}$ and $k_z$ (Table 2). The values obtained from the simulation give high values of adjusted R-square (> 0.99) for all the samples, indicating the high accuracy of the values obtained for $k_{xy}$ and $k_z$. Further, multiple measurements were performed at different points in the sample to eliminate the effect of local variation of defect density on the surface of the sample. No substantial change in thermal conductivity values was observed, and the variation remained within the estimated error values of the quantified thermal conductivity value. Therefore, for the sake of clarity, only one measurement point and analysis have been discussed in the manuscript. The overall value of $\kappa_{tl}$ (in the present case, $Sb_2Te_3$-$AgSbTe_2$) is very low and lies in the range of 0.7-1.0 $Wm^{-1}K^{-1}$ for all the composite samples. These values observed in the case of composite samples are extremely low as compared to the value of thermal conductivity of pristine $Sb_2Te_3$ ($Sb_2Te_3$ > 1.6 $Wm^{-1}K^{-1}$) and presented in table 2. Further, it can be observed from the $k_{xy}$

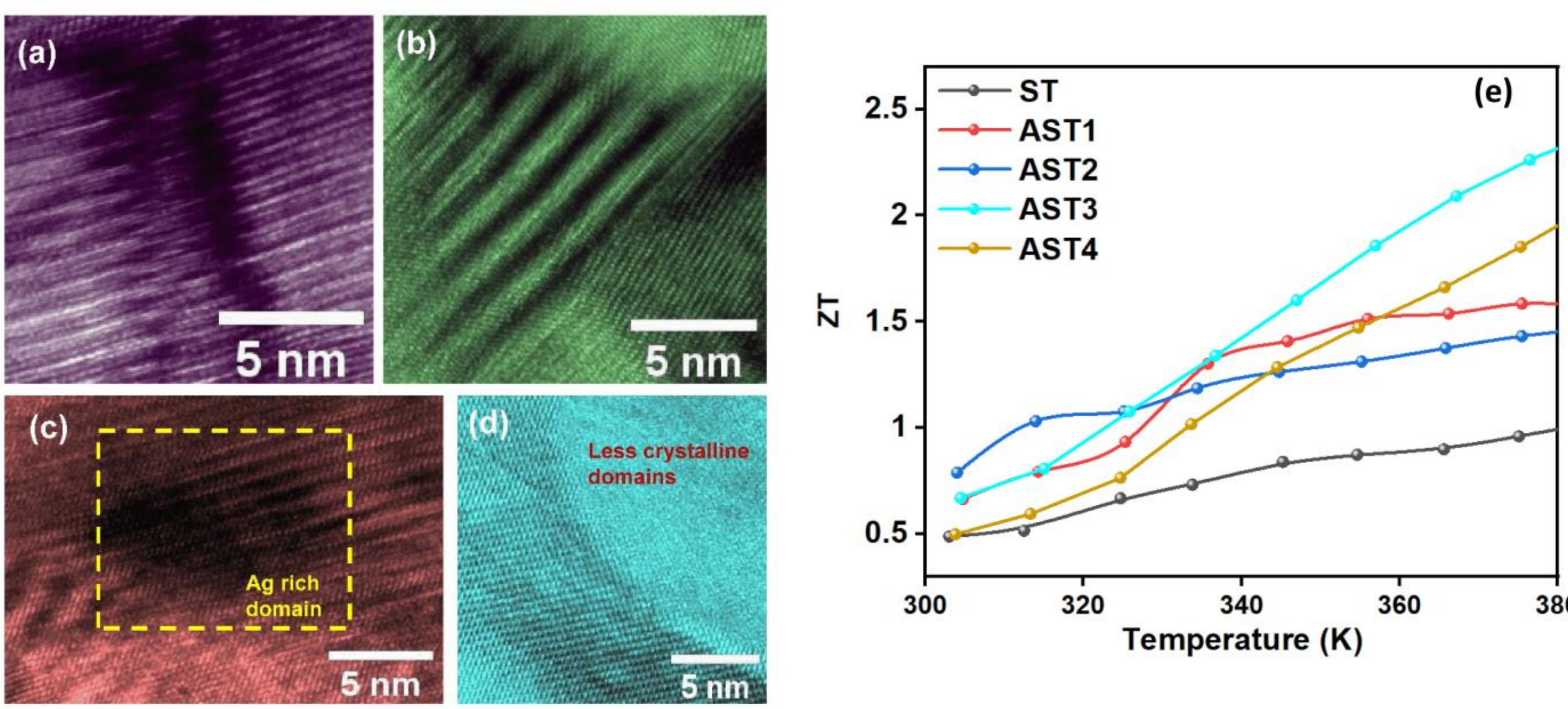

Figure 9 (a-d) shows the HRTEM images corresponding to AST4 showing different kind of defect formations. The dark areas represent Ag rich domains. (e) shows the in plane ZT value of all the samples.

and $k_z$ values that all the composite samples not only demonstrate overall lower values of thermal conductivity as compared to the pristine $Sb_2Te_3$ sample but also offer combined enhanced phonon scattering in the in-plane as well as across the plane directions. In order to validate the large decrease in thermal conductivity value, the microstructural analysis is performed using a field emission scanning electron microscope (FESEM), and the results are presented in Supporting information Fig. S7 . Scanning electron microscopy images indicate that all the samples consist of crystallites with an irregular shape having a size less than 50 nm with well-defined grain boundaries. The size of the crystallites decreases with an increase in Ag content which is also supported by the XRD data. In the case of AST film, a small amount of Ag in the $Sb_2Te_3$ perturb the crystal perfection, and a large number of point and planner defects are generated, as can be seen from Fig. 9 (a-d). As the crystal nucleation usually starts at defects and interfaces due to high interfacial energy and weak bonding, more defects lead to the formation of more nucleation centres. The grain size should be decreased to the same volume to maintain the balance. Besides that, the more considerable electronegativity difference between Ag-Te and Ag-Sb compared to Sb-Te may also increase nucleation density. Higher nuclei concentrations are likely to form with Ag incorporation, producing more grain boundaries and significantly suppressing crystal growth. This may explain the smaller grain size in samples having Ag (AST1 to AST4). Similar grain size reduction is observed for $Sb_2Te_3$ doped with Al, W, and Cr. [28, 29, 47] Previous studies have confirmed that grain boundary interfaces can inhibit phonon propagation, which could result in an exceptional reduction in thermal conductivity. [48, 49]Several nanoscale black domains are also observed in the composite sample, and the density increases with an increase in Ag content, leading to the conclusion that the domains may be Ag-rich. These precipitates are associated with high-density lattice distortions, such as linear dislocation, which indicate that the structural modifications occur locally to relax the strain in the composite samples during the growth process. It is well established that heat-carrying phonons in a material cover a broad spectrum of wavelengths. To lower thermal conductivity values, it is necessary to scatter the phonons of all energies. The presence of atomic-scale distortion in the composite samples restricts the propagation of short-wavelength phonons, while the hetero-interface and the grain boundary can strongly scatter mid to long wavelength phonons. The presence of the distinct boundary and the evolution of two separate phases of $Sb_2Te_3$ and $AgSbTe_2$ are likely to provide a high density of interfaces that result in the scattering of long-wave-length phonons. Thus, considering sources of scattering across all length scales hierarchically from atomic lattice disorder and nanoscale secondary phases to grain boundaries, lattice thermal conductivity is reduced. The reduction in thermal conductivity leads to a significant enhancement in the thermoelectric properties of such composite structures. A high ZT value of 2.26 and 1.85 is achieved in sample AST3 and sample AST4 at 375K which is a highest reported value in AST based TE materials in this temperature range. The values of ZT at these temperatures were calculated assuming a constant thermal conductivity value at high temperature, that is a typical trend for the systems where the thermal conductivity is dominated by phonons, that was widely reported elsewhere. The only exception would be the bipolar heat transport that is not observed in our system up to 380 K. [50, 51]

## Conclusions

In this work, a large improvement in the Seebeck coefficient and reduction in thermal conductivity results in an overall enhancement of the thermoelectric properties of $Sb_2Te_3$-$AgSbTe_2$ thin film nanocomposites. The enhancement in the power factor has been understood by carrying out a detailed study of microstructural and electronic properties and discussed in terms of the band structure modifications, which includes band flattening, band gap elevation, the formation of resonance level, and carrier filtering effect. At the same time, reduced thermal conductivity is explained through phonon scattering from the heterointerfaces, defects, and grain boundary scattering resulting with experimentally measured reduction of in-plane thermal conductivity value by ~ 58% upon the incorporation of the $AgSbTe_2$ phase in the $Sb_2Te_3$ matrix. The significant decline in carrier mobility and charge density certifies a large enhancement in effective mass in the composite sample. Along with a significant enhancement in the power factor value, reduced thermal conductivity values lead to high ZT values at high temperatures. The maximum value of 2.2 value is achieved is achieved at 375 K which is comparable to the value of SnSe.[52] This study firmly demonstrates the possibility of improving TE properties through band-structure modifications and carrier filtering effect for improving the thermoelectric properties by carrying out a controlled synthesis protocol that can be extended to other thermoelectric systems. Further, the study provides newer pathways for enhancing the TE performance of nanocomposites but also demonstrates a novel method of determining anisotropic values of thermal conductivity, which otherwise is strenuous to obtain, especially in the case of complex structures and geometries.

## Experimental details

**Synthesis:** $Sb_2Te_3$-$AgSbTe_2$ thin nanocomposite films were prepared on resistive Si/$SiO_2$ (100) substrate at 523K by radio frequency (RF) magnetron sputtering technique followed by a post-annealing treatment at the same temperature for 30 minutes in an inert atmosphere. Two separate targets of $Sb_2Te_3$ and Ag are used for this study. The RF power for $Sb_2Te_3$ was maintained constant at 40 W, while the Ag content was varied by changing the power of the Ag target. A series of six samples were prepared with RF power ranging from 10 Watt to 20 Watt in the step of two Watt. During deposition, the working pressure was maintained at $2 \times 10^{-3}$ millibar.

## Samples characterization:

The structural and morphological properties of all the prepared samples were characterized using X-ray diffraction (XRD), X-ray photoelectron spectroscopy (XPS), Raman, Field emission scanning electron microscopy (FESEM) measurements, and transmission electron microscopy (TEM).

**XRD**

X-ray diffraction patterns were obtained between 20°-80° at a glancing angle with PANalytical X'Pert PRO X-ray diffractometer with Cu Kα irradiation (λ=0.154 178 nm).

**XPS**

X-ray photoelectron spectroscopy (XPS) measurements were performed using the ESCA+ Omicron nanotechnology with the monochromatic X-ray excitation source having an energy Al Kα line (1486.6 eV). The obtained spectra were calibrated using amorphous carbon (C 1s peak at 284.8 eV) present in the sample.

**Raman**

Raman analysis of all the samples was premeditated on Renishaw invia confocal Raman microscope with a 514 nm laser wavelength at room temperature. The laser power of < 1 mW was used to control sample damage.

**FESEM**

The morphology of the films was characterized using ZEISS EVO 50 Scanning electron microscope,

**Cross-sectional HRTEM**

Cross-sectional HRTEM was performed with a double spherical aberration-corrected, cold field emission gun (FEG) JEOL ARM 200FC, operated at 200 kV. This instrument is equipped with a 100 $mm^2$ large solid angle (covering a solid angle of 0.98 sr) Centurio detector for energy-dispersive X-ray spectroscopy (EDS). The sample was coated with a thin layer of gold (deposited by a Cressington sputter coater) prior to FIB sample preparation to get rid of sample charging. The region of interest for TEM was coated with carbon. The first part of the carbon protection layer was made with e-beam-assisted carbon deposition. This layer was followed by a thicker layer made by ion-beam-assisted carbon deposition. All coarse thinning was performed at 30 kV acceleration voltage for the Ga+ ions. Final thinning was first done at 5 kV and finally at 2 kV on either side of the lamellae to minimize surface damage.

**Kelvin probe force microscopy:**

The KPFM technique was performed using Dimension Icon from Bruker with Pt– Ir coated Silicon tip (SCM-PIT from Bruker, Inc., USA), having a radius of curvature of 30 nm and a resonance frequency of 50 kHz. The bias was applied to the tip, and a constant lift height of 50 nm was maintained for measurement. The surface potential is measured with a scan rate of 0.8 Hz. The work function of the tip was calibrated from the surface potential of the HOPG sample, which was used to calculate the work function of the samples.

**Thermoelectric power factor measurement:**

The temperature-dependent Seebeck coefficient was measured in Linseis LSR-3 system under Helium atmosphere.A detailed description of the system is provided in the supporting information Fig. S8.

.

**Hall Measurements:**

Hall measurements were performed in four probe (Van dar pau configurations) utilizing the Physical Property Measurement System (PPMS, Quantum Design) with the AC transport option at room temperature. The magnetic field was set at zero before measurements, and the AC bridge potentiometer nullified the voltage offset. The measurements were performed by measuring the Hall voltage as a function of the magnetic field B (varied from -7T to +7 T) with a step of 1500 Gauss. After the measurement, the magneto-resistive component was omitted by averaging out the measured Hall coefficients by the following relation

$$R_H = \frac{R_H(+B) + R_H(-B)}{2}$$

**Sample preparation for xSThM using BEXP wedge cut:**

The samples were placed on a stage, which is tilted at an angle (~3-5°) with respect to the horizontal plane. After that, three intersecting Ar-ion beams enter the sample side as shown in Fig.5b. As the stage is tilted at a negative angle, the beam enters the side of the sample and exits the sample surface at a glancing angle, creating a wedge-like geometry. This method thus is termed as beam exit nano-cross-section polishing (BEXP). The BEXP technique for creating wedge cuts not only allows the subsurface analysis but also provides minimum damage to the resulting cut with a nm range of surface roughness due to the glancing angle of the ion beam, the inert nature of the Ar ions, and the finishing polishing step of low-energy Ar-ions. To obtain smooth surface, the samples were placed around 25 µm above the height of shadow mask. The polishing process consists of four steps a) firstly, the guns are warmed up at ~ 1 kV, b) pre-polishing of the samples were performed at ~ 3 kV, c) actual polishing of the surface at ~ 7 kV and finally d) smoothening of the as obtained wedge cut at ~ 1 kV.

**Computation:** First-principles calculations were performed within density functional theory (DFT) using the projected augmented-wave method implemented in the Vienna sab initio simulation package (VASP). The generalized gradient approximation (GGA) of the Perdew-Burke-Brinkerhoff was employed to treat the exchange-correlation interaction. The structures Sb2Te3 and Sb2Te3/AgSbTe2 were properly optimized until the force on each atom became less than 0.0001 eV/ Å. The irreducible Brillouin zone was sampled with a set of (11 11 1) Monkhorst-Pack grid to generate k-points during relaxation, while values of (15 15 1) were used for band structure and density of states calculation.

**Acknowledgments:**

B. R. Mehta acknowledges the support of the Schlumberger Chair Professorship and the project funded by DST (Project No. DST/NM/NS/2018/234(G) and INT/NOR/RCN/ICT/P-04/2018). This work was supported by the Research Council of Norway (Project no. 280788) and grant 197405 for the TEM Gemini Centre, NTNU, Norway. O. Kolosov and K. Agarwal acknowledge support of EU Graphene Flagship Core 3 project, and EPSRC project EP/V00767X/1. Authors are grateful to Bruker UK and Leica Microsystems for the help with the advanced operation of the instrumentation. The authors acknowledge the Nanoscale Research facility, Central Research facility, and the department of physics IIT Delhi for providing necessary facilities. Chandan K Vishwakarma acknowledges Dr. B K Mani for providing computational facilities. MA also acknowledges the Foundation for Polish Science through the International Research Agendas (IRA) Programme co-financed by EU within the Smart Growth Operational Programme (SG OP), grant number MAB/2017/1.

## References:

1. E, B.L., *Cooling, Heating, Generating Power, and Recovering Waste Heat with Thermoelectric Systems.* Science, 2008. **321**(5895): p. 1457-1461.
2. Rowe, D.M., *CRC handbook of thermoelectrics*. 1995, Boca Raton, FL: CRC Press.
3. Zhang, Y., *Thermoelectric Advances to Capture Waste Heat in Automobiles.* ACS Energy Letters, 2018. **3**(7): p. 1523-1524.
4. Jia, Y., et al., *Wearable Thermoelectric Materials and Devices for Self-Powered Electronic Systems.* 2021. **33**(42): p. 2102990.
5. Cook, B.A., et al., *Analysis of Nanostructuring in High Figure-of-Merit Ag1−xPbmSbTe2+m Thermoelectric Materials.* Advanced Functional Materials, 2009. **19**(8): p. 1254-1259.
6. Fang, H.K., et al., *Cubic AgPbmSbTe2+m: Bulk Thermoelectric Materials with High Figure of Merit.* Science, 2004. **303**(5659): p. 818-821.
7. Zhou, M., J.-F. Li, and T. Kita, *Nanostructured AgPbmSbTem+2 System Bulk Materials with Enhanced Thermoelectric Performance.* Journal of the American Chemical Society, 2008. **130**(13): p. 4527-4532.
8. Kanatzidis, M.G., *Nanostructured Thermoelectrics: The New Paradigm?* Chemistry of Materials, 2010. **22**(3): p. 648-659.
9. Quarez, E., et al., *Nanostructuring, Compositional Fluctuations, and Atomic Ordering in the Thermoelectric Materials AgPbmSbTe2+m. The Myth of Solid Solutions.* Journal of the American Chemical Society, 2005. **127**(25): p. 9177-9190.
10. Kim, H.-S., et al., *Large-scale production of (GeTe)x(AgSbTe2)100−x (x=75, 80, 85, 90) with enhanced thermoelectric properties via gas-atomization and spark plasma sintering.* Acta Materialia, 2017. **128**: p. 43-53.
11. Yang, S.H., et al., *Nanostructures in high-performance (GeTe)<sub><i>x</i></sub>(AgSbTe<sub>2</sub>)<sub>100−<i>x</i></sub>thermoelectric materials.* Nanotechnology, 2008. **19**(24): p. 245707-245707.
12. Placheova, S.K., *Thermoelectric figure of merit of the system (GeTe)1−x(AgSbTe2)x.* physica status solidi (a), 1984. **83**(1): p. 349-355.
13. Ye, L.-H., et al., *First-principles study of the electronic, optical, and lattice vibrational properties of ${\text{AgSbTe}}_{2}$.* Physical Review B, 2008. **77**(24): p. 245203-245203.
14. Chen, Y., et al., *Transport properties and valence band feature of high-performance (GeTe)85(AgSbTe2)15thermoelectric materials.* New Journal of Physics, 2014. **16**(1): p. 13057-13057.
15. Zhang, Y., et al., *Defect-Engineering-Stabilized AgSbTe2 with High Thermoelectric Performance.* Advanced Materials, 2022. **n/a**(n/a): p. 2208994-2208994.
16. Hong, M., et al., *Achieving zT > 2 in p-Type AgSbTe2−xSex Alloys via Exploring the Extra Light Valence Band and Introducing Dense Stacking Faults.* Advanced Energy Materials, 2018. **8**(9): p. 1702333-1702333.
17. Morelli, D.T., V. Jovovic, and J.P. Heremans, *Intrinsically Minimal Thermal Conductivity in Cubic $\mathrm{I}\mathrm{\text{\ensuremath{-}}}\mathrm{V}\mathrm{\text{\ensuremath{-}}}{\mathrm{VI}}_{2}$ Semiconductors.* Physical Review Letters, 2008. **101**(3): p. 35901-35901.
18. Han, M.-K., et al., *Lead-Free Thermoelectrics: High Figure of Merit in p-type AgSnmSbTem+2.* 2012. **2**(1): p. 157-161.
19. Chen, Y., et al., *SnTe–AgSbTe2 Thermoelectric Alloys.* 2012. **2**(1): p. 58-62.
20. Robson, A.J., et al., *High-accuracy analysis of nanoscale semiconductor layers using beam-exit ar-ion polishing and scanning probe microscopy.* ACS applied materials & interfaces, 2013. **5**(8): p. 3241-3245.
21. Bosse, J.L., et al., *Nanomechanical morphology of amorphous, transition, and crystalline domains in phase change memory thin films.* Applied Surface Science, 2014. **314**: p. 151-157.

22. Maiti, A., et al., *Quantifying anisotropic thermal transport in two-dimensional perovskite ${{\mathrm{PEA}}_{2}{\mathrm{PbI}}_{4})$ through cross-sectional scanning thermal microscopy.* Physical Review Materials, 2023. **7**(2): p. 23801-23801.
23. Grishin, I., B.D. Huey, and O.V. Kolosov, *Three-Dimensional Nanomechanical Mapping of Amorphous and Crystalline Phase Transitions in Phase-Change Materials.* ACS Applied Materials & Interfaces, 2013. **5**(21): p. 11441-11445.
24. Kolosov, O.V., I. Grishin, and R. Jones, *Material sensitive scanning probe microscopy of subsurface semiconductor nanostructures via beam exit Ar ion polishing.* Nanotechnology, 2011. **22**(18): p. 185702.
25. Xu, J., et al., *Crystallization and C-RAM application of Ag-doped Sb2Te3 material.* Materials Science and Engineering: B, 2006. **127**(2): p. 228-232.
26. Yi-Ming, C. and P.C. Kuo, *Effect of Ag or Cu doping on erasable phase-change Sb-Te thin films.* IEEE Transactions on Magnetics, 1998. **34**(2): p. 432-434.
27. Hwang, S., et al., *Ultra-low Energy Phase Change Memory with Improved Thermal Stability by Tailoring the Local Structure through Ag Doping.* ACS Applied Materials & Interfaces, 2020. **12**(33): p. 37285-37294.
28. Xia, M., et al., *Aluminum-Centered Tetrahedron-Octahedron Transition in Advancing Al-Sb-Te Phase Change Properties.* Scientific Reports, 2015. **5**(1): p. 8548.
29. Collins-McIntyre, L.J., et al., *Structural, electronic, and magnetic investigation of magnetic ordering in MBE-grown Cr x Sb 2−x Te 3 thin films.* EPL (Europhysics Letters), 2016. **115**(2): p. 27006-27006.
30. *The Nature of the Chemical Bond and the Structure of Molecules and Crystals.* Nature, 1941. **148**(3762): p. 677-677.
31. Navrátil, J., et al., *Behavior of Ag Admixtures in Sb2Te3and Bi2Te3Single Crystals.* Journal of Solid State Chemistry, 1998. **140**(1): p. 29-37.
32. Medlin, D.L. and G.J. Snyder, *Atomic-Scale Interfacial Structure in Rock Salt and Tetradymite Chalcogenide Thermoelectric Materials.* JOM, 2013. **65**(3): p. 390-400.
33. Sugar, J.D. and D.L. Medlin, *Solid-state precipitation of stable and metastable layered compounds in thermoelectric AgSbTe2.* Journal of Materials Science, 2011. **46**(6): p. 1668-1679.
34. Wolfe, R., J.H. Wernick, and S.E. Haszko, *Anomalous Hall Effect in AgSbTe2.* Journal of Applied Physics, 2004. **31**(11): p. 1959-1964.
35. Jovovic, V. and J.P. Heremans, *Measurements of the energy band gap and valence band structure of ${\text{AgSbTe}}_{2}$.* Physical Review B, 2008. **77**(24): p. 245204.
36. Klichová, I., et al., *Characterization of Ag-doped Sb1.5Bi0.5Te3 single crystals.* Radiation Effects and Defects in Solids, 1998. **145**(3): p. 245-262.
37. Agarwal, K. and B.R. Mehta, *Structural, electrical, and thermoelectric properties of bismuth telluride: Silicon/carbon nanocomposites thin films.* Journal of Applied Physics, 2014. **116**(8).
38. Wiendlocha, B., et al., *Residual resistivity as an independent indicator of resonant levels in semiconductors.* Materials Horizons, 2021. **8**(6): p. 1735-1743.
39. Ioffe, A.F. and C. Wert, *Physics of Semiconductors.* Journal of The Electrochemical Society, 1962. **109**(2): p. 43C.
40. Pei, Y., et al., *Low effective mass leading to high thermoelectric performance.* Energy & Environmental Science, 2012. **5**(7): p. 7963-7969.
41. Kane, E.O., *Band structure of indium antimonide.* Journal of Physics and Chemistry of Solids, 1957. **1**(4): p. 249-261.
42. Wiendlocha, B., *Thermopower of thermoelectric materials with resonant levels: PbTe:Tl versus PbTe:Na and ${\mathrm{Cu}}_{1\ensuremath{-}x}{\mathrm{Ni}}_{x}$.* Physical Review B, 2018. **97**(20): p. 205203.
43. Ghosh, A., et al., *Modifying the Thermoelectric Transport of Sb2Te3 Thin Films via the Carrier Filtering Effect by Incorporating Size-Selected Gold Nanoparticles.* ACS Applied Materials & Interfaces, 2021. **13**(11): p. 13226-13234.
44. Ahmad, M., et al., *The nature of 2D:3D SnS:Bi2Te3 interface and its effect on enhanced electrical and thermoelectric properties.* Journal of Alloys and Compounds, 2020. **847**: p. 156233.
45. Muzychka, Y.S., *Spreading Resistance in Compound Orthotropic Flux Tubes and Channels with Interfacial Resistance.* Journal of Thermophysics and Heat Transfer, 2014. **28**(2): p. 313-319.

46. Yovanovich, M.M., Y.S. Muzychka, and J.R. Culham, *Spreading Resistance of Isoflux Rectangles and Strips on Compound Flux Channels.* Journal of Thermophysics and Heat Transfer, 1999. **13**(4): p. 495-500.
47. Wang, G., et al., *Crystallization behaviors of W-doped Sb3Te phase change films.* Vacuum, 2015. **121**: p. 142-146.
48. Chen, Z., X. Zhang, and Y. Pei, *Manipulation of Phonon Transport in Thermoelectrics.* 2018. **30**(17): p. 1705617.
49. Kim, W., *Strategies for engineering phonon transport in thermoelectrics.* Journal of Materials Chemistry C, 2015. **3**(40): p. 10336-10348.
50. Yu, Z., et al., *Phase-dependent thermal conductivity of electrodeposited antimony telluride films.* Journal of Materials Chemistry C, 2018. **6**(13): p. 3410-3416.
51. Ferrer-Argemi, L., et al., *Silver content dependent thermal conductivity and thermoelectric properties of electrodeposited antimony telluride thin films.* Scientific Reports, 2019. **9**(1): p. 9242-9242.
52. Zhao, L.-D., et al., *Ultrahigh power factor and thermoelectric performance in hole-doped single-crystal SnSe.* 2016. **351**(6269): p. 141-144.